\documentclass[conference]{IEEEtran}
\IEEEoverridecommandlockouts
\usepackage[T1]{fontenc}
\usepackage{cite}
\usepackage{amsmath,amssymb,amsfonts}
\usepackage{algorithmic}
\usepackage{graphicx}
\usepackage{comment}
\usepackage{textcomp}
\usepackage{xcolor}
\usepackage{tabu} 
\usepackage{booktabs}
\usepackage{siunitx}
\usepackage{soul}
\usepackage{amsfonts}
\def\BibTeX{{\rm B\kern-.05em{\sc i\kern-.025em b}\kern-.08em
    T\kern-.1667em\lower.7ex\hbox{E}\kern-.125emX}}
\begin{document}

\title{Behavioral Biometrics for Automatic Detection of User Familiarity in VR}


\author{
    \IEEEauthorblockN{Numan Zafar, Priyo Ranjan Kundu Prosun, Shafique Ahmad Chaudhry}
    \IEEEauthorblockA{Clarkson University, Potsdam, NY, USA
    \\\texttt{\{zafarn,prosunp,schaudhr\}@clarkson.edu}}
}
\maketitle

\begin{abstract}

As virtual reality (VR) devices become increasingly integrated into everyday settings, a growing number of users without prior experience will engage with VR systems. Automatically detecting a user’s familiarity with VR as an interaction medium enables real-time, adaptive training and interface adjustments, minimizing user frustration and improving task performance. In this study, we explore the automatic detection of VR familiarity by analyzing hand movement patterns during a passcode-based door-opening task, which is a well-known interaction in collaborative virtual environments such as meeting rooms, offices, and healthcare spaces. While novice users may lack prior VR experience, they are likely to be familiar with analogous real-world tasks involving keypad entry. We conducted a pilot study with 26 participants, evenly split between experienced and inexperienced VR users, who performed tasks using both controller-based and hand-tracking interactions. Our approach uses state-of-the-art deep classifiers for automatic VR familiarity detection, achieving the highest accuracies of 92.05\% and 83.42\% for hand-tracking and controller-based interactions, respectively. In the cross-device evaluation, where classifiers trained on controller data were tested using hand-tracking data, the model achieved an accuracy of 78.89\%. The integration of both modalities in the mixed-device evaluation obtained an accuracy of 94.19\%. Our results underline the promise of using hand movement biometrics for the real-time detection of user familiarity in critical VR applications, paving the way for personalized and adaptive VR experiences.

\end{abstract}


\section{Introduction}

Virtual reality (VR) technology is becoming an integral part of daily life, offering immersive experiences that merge the boundaries between physical and virtual environments. The adoption of VR is expanding across industries and consumer applications, such as education~\cite{gao2021digital}, retail~\cite{xi2021shopping}, personal banking, and finance~\cite{mathis2022can}. VR in the therapeutic domain is used to treat anxiety~\cite{chard2022virtual} and phobias~\cite{ramalhoto2024phobeevr}. In fitness programs, VR elevates exercise engagement~\cite{hayes2024more}, while VR provides a safe and controlled environment for military training~\cite{de2020use, rettinger2022defuse}, and emergency simulations~\cite{munsinger2023virtual}. As VR has become more prevalent, the number of novice users is anticipated to increase. However, a lack of familiarity with VR can lead to frustration and disengagement due to the challenges associated with navigating and interacting within virtual environments, which may seem overwhelming for users with limited experience. Therefore, understanding a user's initial level is essential to ensure that VR can adapt to different user levels.

Studies have shown that user experience with different interaction devices and immersion levels affects user behavior in VR~\cite{speicher2018xd, abdi2019using}. The evolution of user interactions in VR across different timescales is becoming important in behavior-based biometric authentication~\cite{liebers2023exploring, miller2022temporal}. Li et al.~\cite{li2024evaluating} were able to identify users with no prior VR experience based on their motion pattern using hand tracking and achieved 88\% accuracy. However, there is still a lack of automated methods to detect the familiarity level with different input modalities.

In this study, we provide a novel approach for automating the detection of user familiarity by analyzing hand movements across diverse input modalities in VR. We modified Li et al.'s task, where users open a virtual door by entering a four-digit passcode, a familiar action regardless of their VR experience. The selected task of inputting the passcode is significant in underscoring the need for adaptive security protocols~\cite{jones2021literature, stephenson2022sok}. In attempting to access secure spaces in VR, user unfamiliarity may lead to incorrect or unintended actions, such as pressing incorrect keys or making imprecise gestures that can result in lockouts by VR authentication mechanisms. To explore the effects of the different modalities, users interact with both hand tracking and controllers. VR doors serve as common entry points into collaborative virtual spaces, such as a virtual office or a shared conference room, where a passcode is needed for access. Detecting prior VR experiences early through interactions with the VR door could enable adaptive assistance before users engage in more complex tasks. In our work, we capture the user's finger movements during the entry of a PIN code sequence in VR to unlock a door. We use these data to train the deep learning classifiers to infer the user's VR familiarity level. As a proxy for familiarity, we use a 1-5 scale for self-rated experience. The study demonstrates that integrating multiple modalities, such as hand tracking and controller-based interaction, significantly enhances familiarity detection and achieves an accuracy of up to 94\%. To the best of our knowledge, this is the first study to automatically detect VR familiarity using multi-modal behavioral biometrics, enabling cross-device generalization. This work lays the groundwork for VR systems that can automatically recognize a user's experience level and integrate such detection for personalized, real-world VR applications.

\section{Related Work}

Recent research in VR has increasingly focused on behavior-based user modeling for identification and authentication using behavioral and biometric signals. For instance, the BehaVR framework integrates gaze direction, facial expressions, and motion data to recognize users without explicit login~\cite{jarin2023behavr}. George et al.~\cite{george2017seamless} investigated how traditional authentication methods such as PINs and Android unlock patterns translate to VR, finding usability challenges and susceptibility to observation attacks. Pfeuffer et al.~\cite{pfeuffer2019behavioural} found that users can be identified based on coordination of their head, hand, and eye movements during natural VR tasks. Similarly, Yu et al.~\cite{yu2016exploration} showed that conventional PIN and pattern lock techniques become vulnerable in immersive environments due to the exposed input actions. Wierzbowski et al.~\cite{wierzbowski2022behavioural} demonstrated that users can be identified based on viewing patterns of 360-degree videos, by analyzing eye tracking and head orientation, without any interactive task. Furthermore, Liebers et al.~\cite{liebers2023exploring} conducted an eight-week remote study to examine the long-term stability of motion-based biometrics in VR authentication and observed that naturally evolving behavior can undermine its effectiveness over time. Ajit et al.~\cite{ajit2019combining} proposed a biometric system by combining pairwise trajectory features of VR headsets and controllers to enable seamless and continuous user identification. However, these studies primarily focus on VR authentication, leaving a significant gap in understanding how user familiarity, experience, and behavior influence performance across diverse VR tasks.

To address this gap, some research has investigated user familiarity within specific VR tasks. Chan et al.~\cite{chan2021systematic} reviewed uses of eye and hand tracking to assess technical skills in neurosurgery training, while eye-tracking has been used to classify soccer goalkeepers' expertise in realistic scenarios using machine learning~\cite{hosp2020eye}.
Scholz et al.~\cite{alinaghi2022can} demonstrated that gaze patterns during wayfinding can reveal spatial familiarity based on prior experience. Castillon et al.~\cite{castillon2024automatically} detected a user's internal sense of familiarity using eye-gaze features in VR, demonstrating potential for adaptive tutoring systems. In gaming, Järvelä et al.~\cite{gerling2011measuring} showed that transitioning to an unfamiliar input modality affects usability, even if the overall experience remains positive. At the system level, user perception has been leveraged to improve immersive rendering in real time through deep learning-based visual prediction models~\cite{an2023development}. Asish et al.~\cite{asish2023detecting} integrated EEG and gaze data to detect cognitive distraction in educational VR using CNN-LSTM and Random Forest classifiers. The work of Li et al. \cite{li2024evaluating} investigated user familiarity using hand tracking in a virtual keypad task by analyzing hand trajectories using deep learning models, highlighting the potential for motion-based familiarity detection in controlled environments. 
Unlike these works, our study addresses this gap by examining how prior VR experience influences behavior across different input methods and by evaluating classifiers that can detect user familiarity and generalize across these modalities.

\begin{figure}[t]
    \centering
        \includegraphics[width=\linewidth]{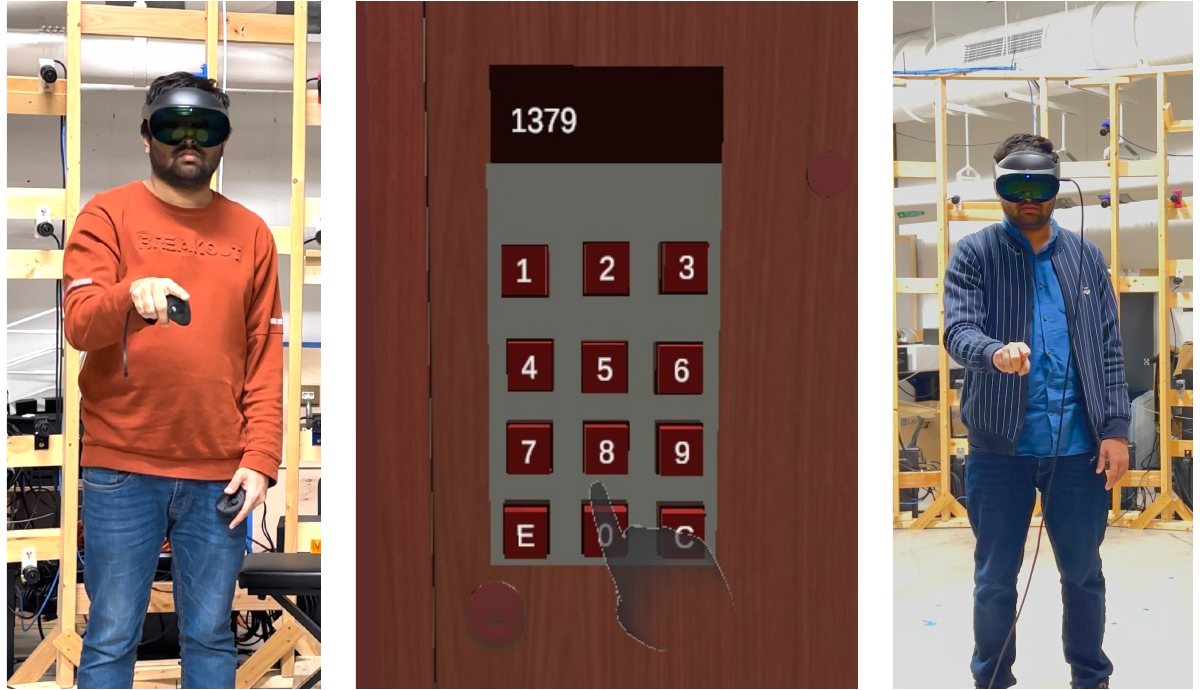}
        \caption{Participants interacting with a VR door unlocking application by inputting code 1379. The left image depicts a participant using a controller, whereas the right uses hand-tracking.}
        \label{fig:interaction}
\end{figure}

\section{Data Collection}

\begin{figure*}[t]
    \centering
        \includegraphics[width=0.843\linewidth]{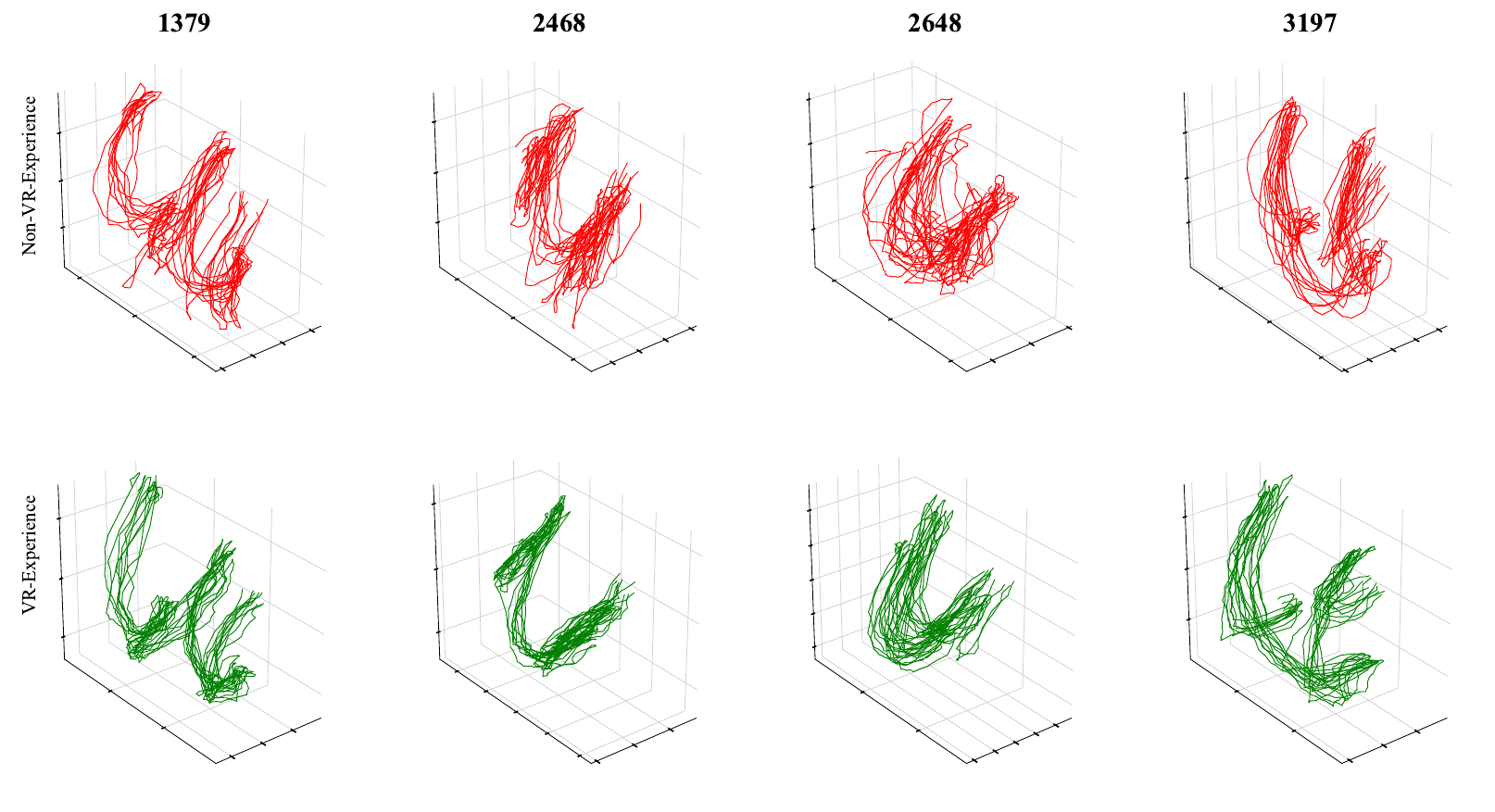}
        \caption{Ten trial right-hand movement trajectories for the four passcodes using hand tracking-based interaction, comparing participants with no VR experience (top) and VR experience (bottom).}
        \label{fig:hand_traj}
\end{figure*}

We utilize the VR door unlock application developed by Li et al.~\cite{li2024evaluating} and modify it to support both hand tracking and controller-based inputs. We recruit 26 participants (\(M_{age}\) 28.7 $\pm$ 5.4 years), and collect data using a Meta Quest Pro device. Prior to the VR task, each participant completes a demographic questionnaire, including a self-rating from 1 (no experience) to 5 (experience) of their prior VR experience. We ensure an equal split of 13 participants with no prior VR experience and 13 with VR experience. Table~\ref{tab:demographics} summarizes the participant demographics and experience levels.

\begin{table}[h]
    \centering
    \caption{Summary of Demographics of our Dataset}
    \label{tab:demographics}
    \resizebox{\columnwidth}{!}{%
        \begin{tabular}{|l|c|c|c|}
        \hline
        \textbf{Category} & \textbf{Count} & \textbf{Age $\pm$ STD} & \textbf{Exp $\pm$ STD} \\ 
        \hline \hline
        Number of Participants & 26 &28.69 $\pm$ 5.35 & 2.92 $\pm$ 1.82\\ 
        \hline
        Number of female Participants & 5 & 26.20 $\pm$ 2.48 & 2.60 $\pm$ 1.96\\ 
        \hline
        Number of male Participants & 21 & 29.29 $\pm$ 5.67 & 3.00 $\pm$ 1.77\\ 
        \hline
        Number of Participants with no VR experience & 13 & 30.38 $\pm$ 4.53 & 1.15 $\pm$ 0.36\\ 
        \hline
        Number of Participants with VR experience & 13 &27.00 $\pm$ 5.56 & 4.69 $\pm$ 0.46 \\ 
        \hline
    \end{tabular}
    }
\end{table}

\begin{figure*}[h]
    \centering
        \includegraphics[width=0.843\linewidth]{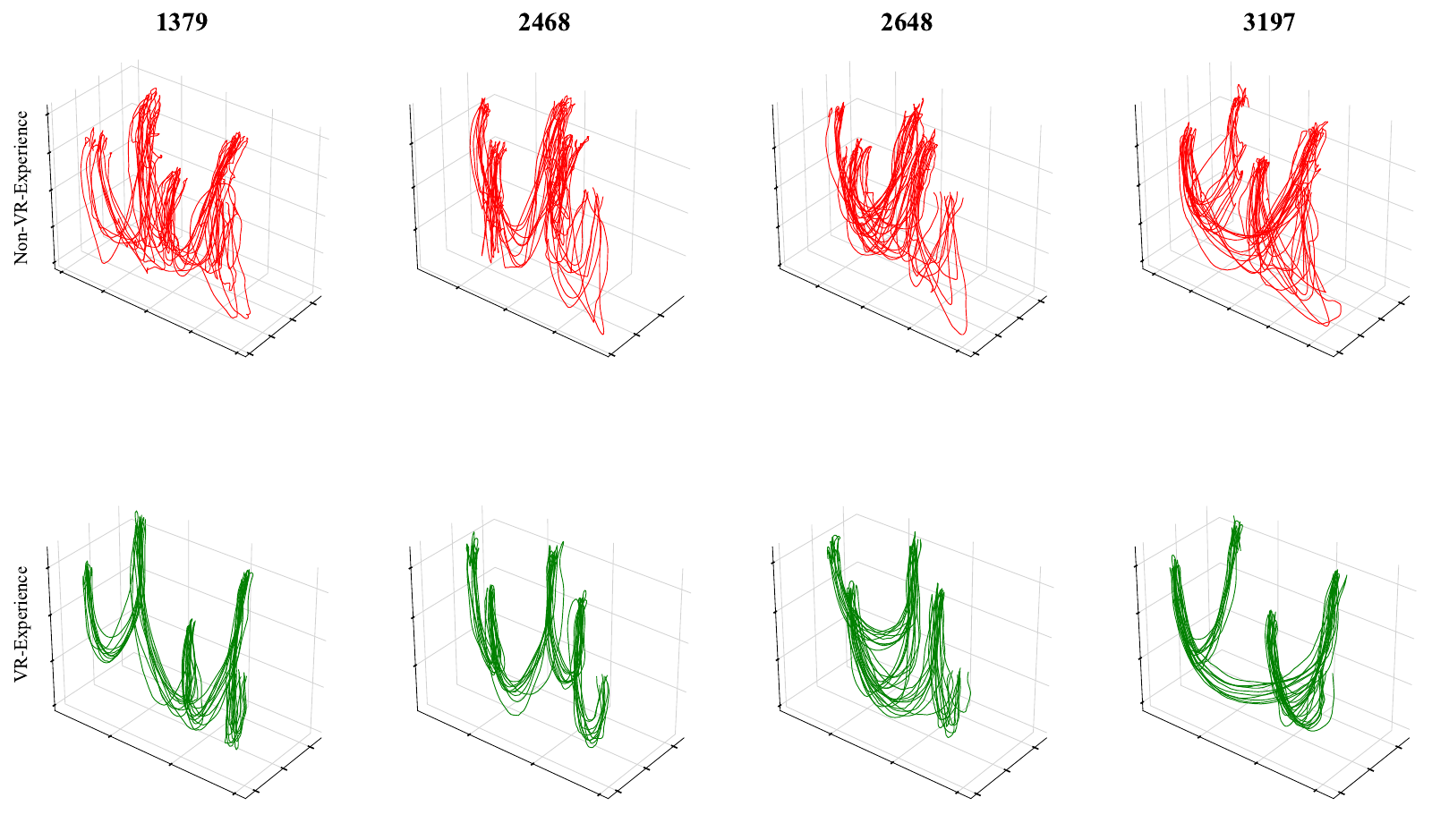}
        \caption{Ten trial right-hand movement trajectories for the four passcodes using controller-based interaction}
        \label{fig:cont_traj}
\end{figure*}

During the experimental task, participants are instructed to enter a four-digit PIN to open a virtual door, as shown in Figure~\ref{fig:interaction}. We use four PIN combinations: 1379, 2468, 2648, and 3197. Each PIN is attempted 10 times, resulting in 40 trials per input modality for each participant. All participants perform the task using both hand tracking and controller-based input, yielding a total of 2080 trials (26 participants × 40 trials × 2 modalities). To mitigate the impact of prior interactions, data collection is conducted in two sessions separated by at least one month. In the first session, half of the participants use hand tracking and the other half controller. In the second session, the input modalities are switched. PINs are presented sequentially for hand tracking and in a randomized order for controller-based interaction. We collect 3D positional and orientation data from the headset, hand tracking, and controller at 72 fps to enable a comparative analysis of interaction behaviors across input modalities. Figures~\ref{fig:hand_traj} and~\ref{fig:cont_traj} illustrate the 3D motion trajectories of participants with (green) and without (red) prior VR experience during the task using hand tracking and controller-based interactions, respectively. Participants without prior VR exposure demonstrate greater variability and inconsistent movement patterns compared to VR-experienced participants, suggesting a lack of spatial coordination and familiarity with the interaction.

\section{Experiment}
We evaluate VR familiarity detection by training classifiers using sliding windows derived from the participants' dominant-hand 3D-positional trajectories in the dataset. We experiment with window lengths of 50, 60, 70, 80, 90, 100, 110, and 120 frames using a fixed step size of 1.

\subsection{Classifiers}

We evaluate several state-of-the-art classifiers to detect VR familiarity. For each window length and PIN combination, we train Long Short-Term Memory (LSTM)~\cite{zebin2018human},  Multi Channel Deep Convolutional Neural Network (MCDCNN)~\cite{ismail2019deep}, Multi-scale Convolutional Neural Network (MCNN)~\cite{cui2016multi}, Time Le-Net (TLENet)~\cite{le2016data}, Transformer Encoder~\cite{vaswani2017attention}, Multilayer Perceptron (MLP), Fully Convolutional Network (FCN)~\cite{wang2017time}, and InceptionTime~\cite{ismail2020inceptiontime}. We find that MLP, FCN, and InceptionTime provide the best performance in terms of accuracy and robustness.

\subsubsection{Multilayer Perceptron} The MLP is a simple feed-forward neural network implemented in PyTorch. It consists of an input layer, two hidden layers, and an output layer. Each hidden layer has a dimensionality equal to half the input size, and uses a ReLU activation function~\cite{agarap2018deep}. The final output layer uses a softmax activation function to produce class probabilities.

\subsubsection{Fully Convolutional Network}

The FCN architecture proposed by Wang et al.~\cite{wang2017time} consists of three convolutional blocks. Each block contains three 1D convolutional layers \{128, 256, 128\} and a 1D kernel \{8, 5, 3\}, followed by batch normalization layers~\cite{ioffe2015batch} to stabilize learning, and a ReLU activation. After the three blocks, we apply a Global Average Pooling (GAP) layer to reduce the feature map dimensionality and prevent overfitting. Finally, the output is passed through a softmax layer to generate class probabilities.



\subsubsection{InceptionTime} 

We use the InceptionTime deep neural network proposed by Fawaz et al.~\cite{ismail2020inceptiontime}, which consists of a sequence of inception modules designed to capture multiscale temporal features. Each module includes a bottleneck layer for dimensionality reduction, followed by three parallel 1D convolutional layers with 1D kernels, and a max-pooling branch with a 1D convolutional layer. The outputs from all branches are concatenated, followed by batch normalization and a ReLU activation. Residual connections are added across inception modules to facilitate gradient flow, followed by a Global Average Pooling (GAP), and a final softmax layer is used to generate class probabilities.

\subsection{Loss Function}
To enhance the performance, models are trained by minimizing a cross-entropy~(\ref{eq1}) loss with label smoothing~(\ref{eq2}):

\begin{equation}
\mathcal{L} = -\frac{1}{|W|} \sum_{w \in W} \sum_{i=1}^{C} q_i^{(w)} \log(p_i^{(w)})
\tag{1}
\label{eq1}
\end{equation}

In Equation~\ref{eq1}, \( \mathcal{L} \) is the average loss over all windows \( w \in W \), \( p_i^{(w)} \) is the softmax probability for class \( i \), and \( q_i^{(w)} \) is the smoothed label for the class. The ground truth label \( y_w \) is set to 0 for users with no prior VR experience and 1 for users with experience.

\begin{equation}
q_i =
\begin{cases}
1 - \varepsilon & \text{if } i = y_w \\
\frac{\varepsilon}{C - 1} & \text{if } i \ne y_w
\end{cases}
\tag{2}
\label{eq2}
\end{equation}

In Equation~\ref{eq2}, \( C = 2 \) represents the number of classes, \( \varepsilon \) is the label-smoothing factor, and \( q_i^{(w)} \) is the smoothed label for the w-th window. To optimize the model parameters, we use the Adam optimizer.

\subsection{Experimental Settings}

We evaluate the VR familiarity detection across four classification scenarios to investigate the model's ability to generalize across various devices and combinations of input data. In each scenario, we employ all ten trials per PIN for each participant and conduct training/testing splits at the participant level to ensure that the model evaluates unseen users. The scenarios are defined as follows:

\subsubsection{Controller-Based Classification} 
The model trains and evaluates using only controller-based interaction data. We randomly select 80\% of the participants (10 experienced, 10 inexperienced) for training and evaluate the remaining 20\% of the participants (3 experienced, 3 inexperienced). This scenario assesses the classifier's ability to distinguish familiarity using the controller motion data.

\subsubsection{Hand-Tracking Based Classification} 
The model trains and evaluates using only hand-tracking-based interaction. Similarly, we select the 80\% of the participants for training and use the remaining 20\% for testing. This evaluates performance using hand-tracking motion data.

\subsubsection{Cross-Device Classification} 
In this scenario, the model is trained using controller-based data and evaluated using hand-tracking data. This setting assesses how well a familiarity classifier trained on one input modality can generalize to a different input device without re-training. 

\subsubsection{Mixed-Device Classification} 
We combine both controller and hand-tracking data for training the models. We mix data from both modalities for 80\% of participants used in training, while testing on the remaining 20\% participants. This approach assesses whether training on a combined modality enhances the overall detection accuracy and robustness across various interaction modes.

\subsection{Implementation Details}
We trained the model on a 12-core AMD Ryzen 9 5900X CPU at 3.7 GHz and an NVIDIA RTX 4090 GPU. All models were trained for 1000 epochs. We trained a total of 384 models, derived from the combination of 4 PINs, 3 classifiers, 8 window sizes, and 4 experimental settings.

\section{Results}

\begin{figure}[htbp]
    \centering
        \includegraphics[width=\linewidth]{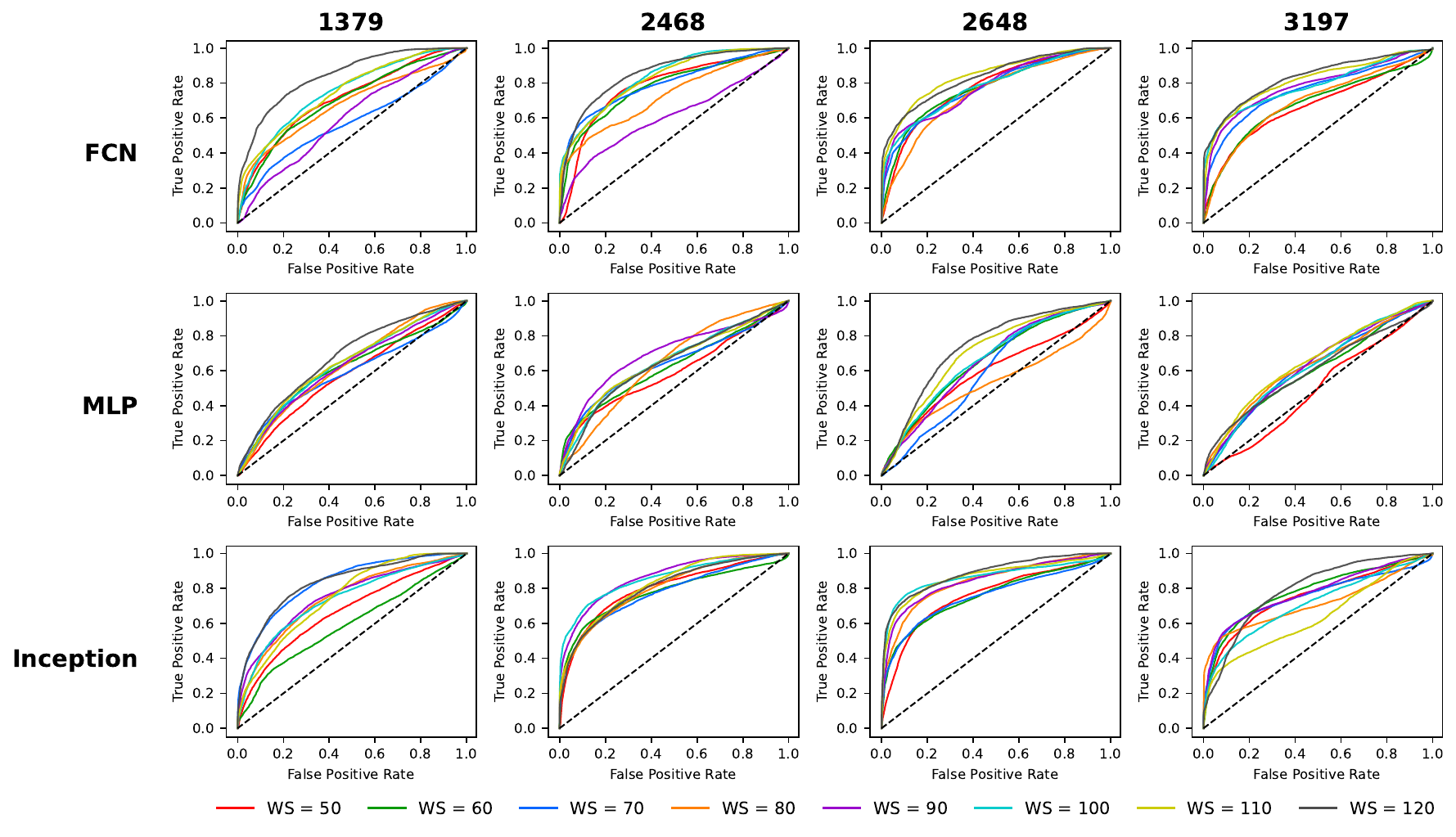}
        \caption{ROC curves show the performance of controller-based classification using the FCN, MLP, and InceptionTime models across different passcodes.}
        \label{fig:cont}
\end{figure}

\begin{figure}[htbp]
    \centering
        \includegraphics[width=\linewidth]{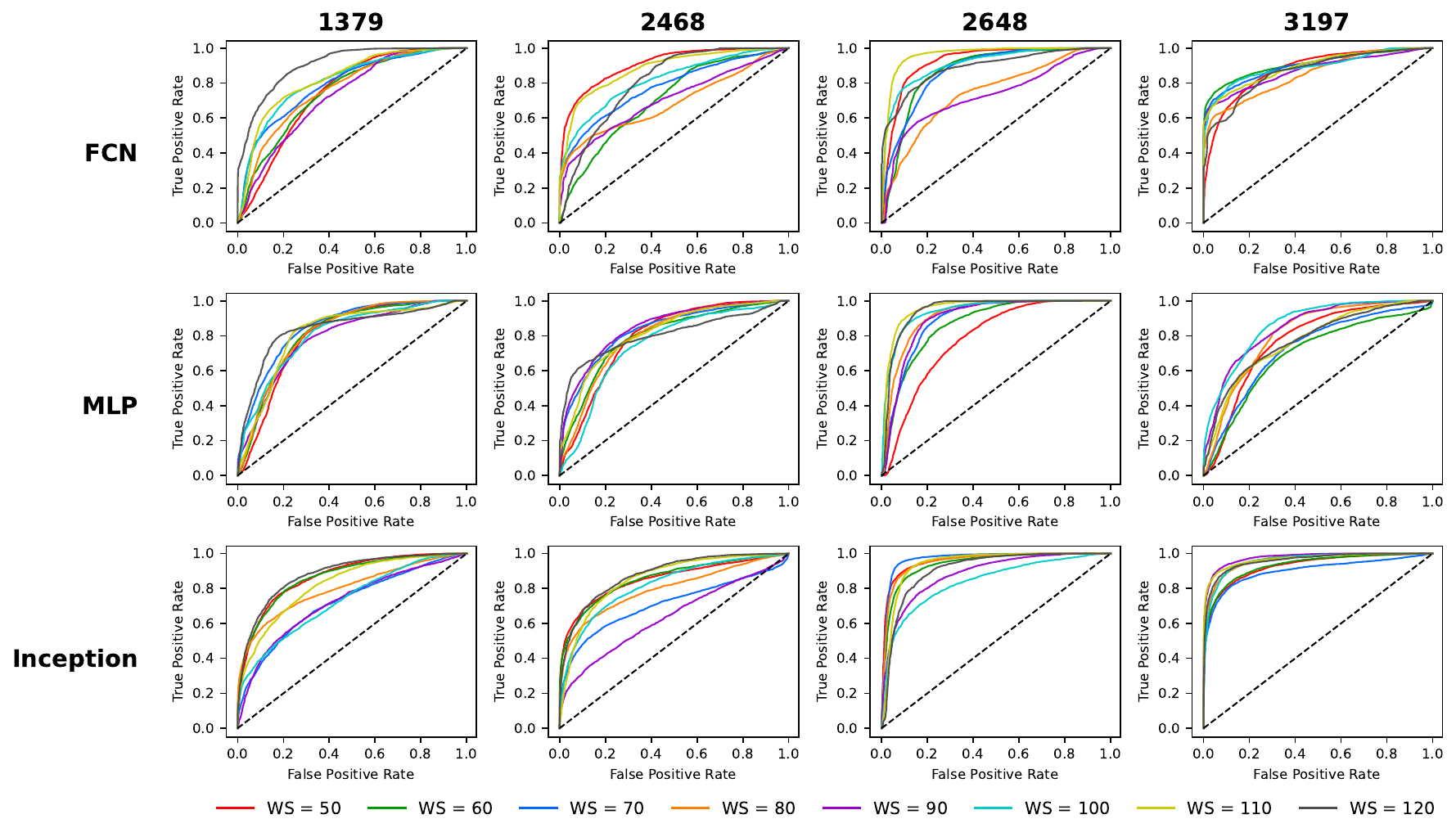}
        \caption{ROC curves for Hand Tracking-based classification using the FCN, MLP, and InceptionTime models across different passcodes.}
        \label{fig:hand}
\end{figure}

\begin{figure}[htbp]
    \centering
        \includegraphics[width=\linewidth]{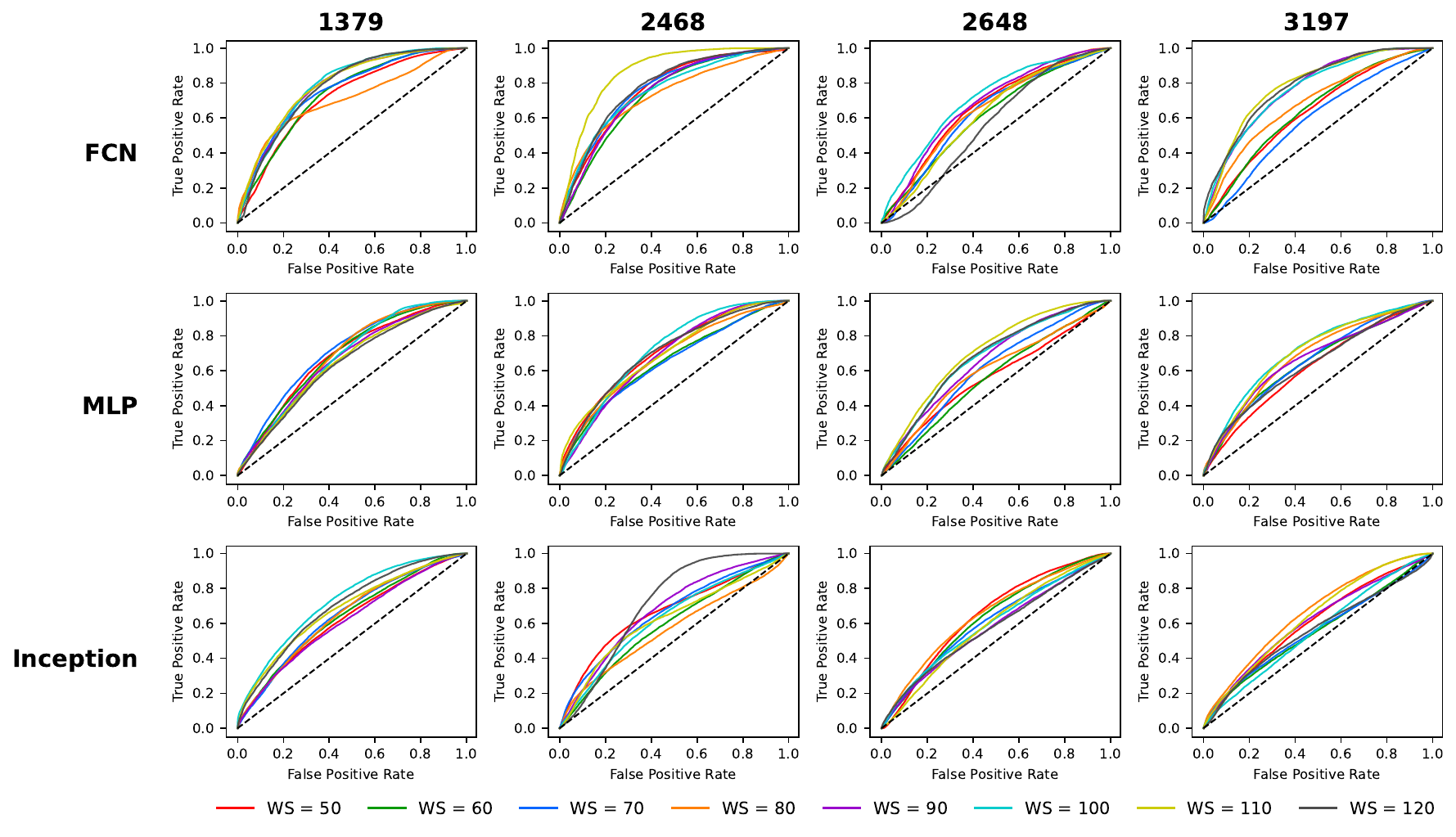}
        \caption{ROC curves for Cross-Device classification using the FCN, MLP, and InceptionTime models across different passcodes.}
        \label{fig:dev}
\end{figure}

\begin{figure}[htbp]
    \centering
        \includegraphics[width=\linewidth]{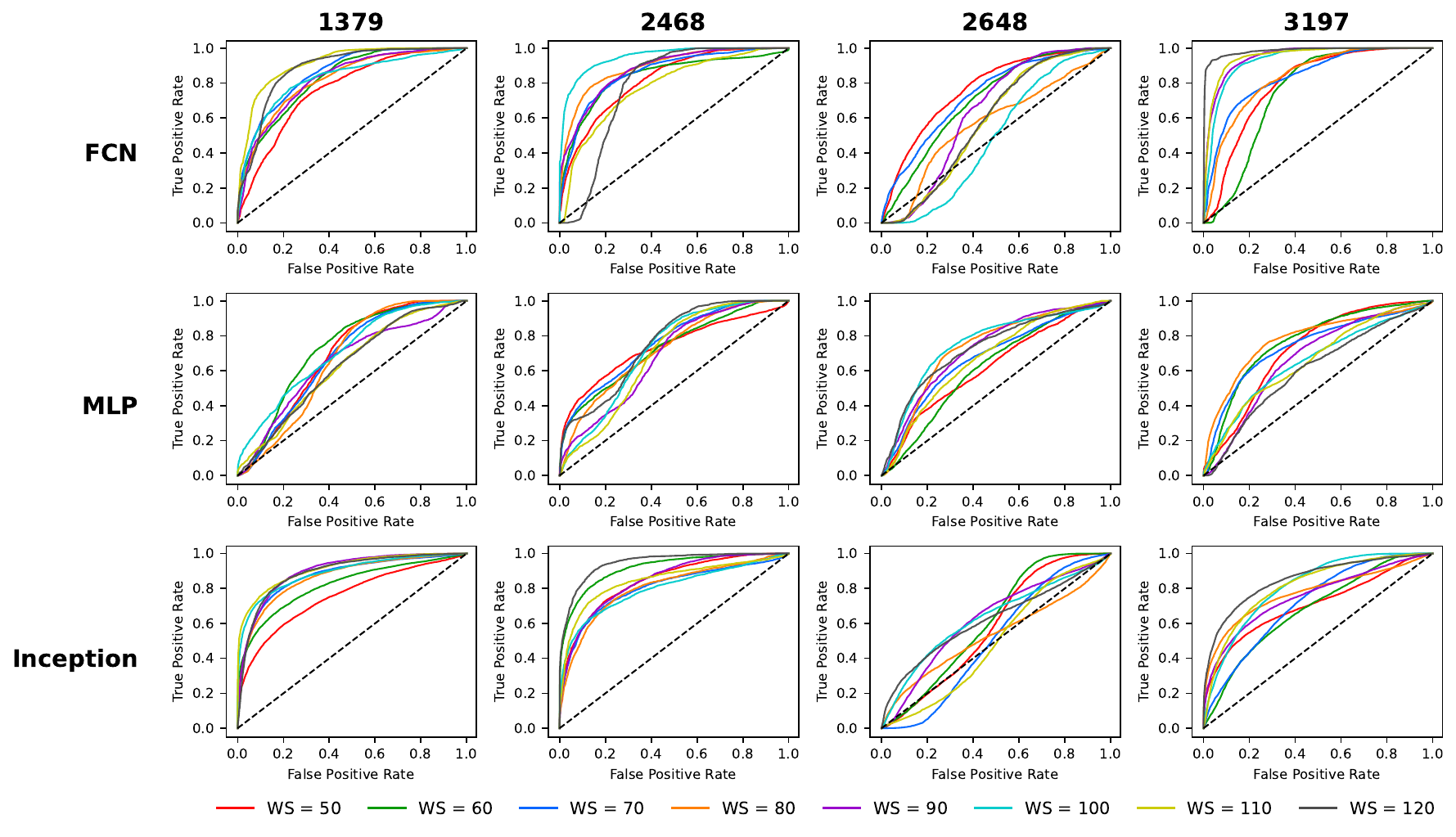}
        \caption{ROC curves for Mixed-Device classification using the FCN, MLP, and InceptionTime models across different passcodes.}
        \label{fig:mix}
\end{figure}

Tables~\ref{tab:cont_acc}--\ref{tab:mix_acc} present the experimental results obtained by employing three neural network classifiers with sliding window sizes ranging from 50 to 120, increasing by a step size of 10 across various PIN combinations. Figures~\ref{fig:cont}--\ref{fig:mix} show the area under the ROC curve (AUC) that evaluates the performance of each model. Each plot contains curves indicating the results for the specific sliding window sizes. The abbreviation 'WS' in the first row indicates the window size values utilized, while the rows represent the classifier types and PIN combinations.

\setlength{\tabcolsep}{3pt}
\begin{table}[htb]
\centering
\caption{Controller Classification Accuracies}
\label{tab:cont_acc}
\resizebox{\columnwidth}{!}{%
  \begin{tabu}{c || cccccccc}
  	\toprule
       WS & 50 & 60 & 70  & 80 & 90 & 100 & 110 & 120 \\ 
      	\hline \hline 
        MLP 1379  & 0.5762 & 0.6199 & 0.6207 & 0.6089 & 0.6147 & 0.6210 & 0.6311 & 0.6375 \\
        FCN 1379  & 0.6593 & 0.6402 & 0.5899  & 0.6647  & 0.5789  & 0.652  & 0.6837  & 0.7639  \\
        INCEPTION 1379  & 0.6430 & 0.6148 & 0.6642 & 0.6877 &0.6715  & 0.7019 & 0.6311 & \textbf{0.7663} \\
        \hline 
        MLP 3197  & 0.5331 & 0.5760 & 0.6030 & 0.6166 & 0.5979 & 0.6047 & 0.6201 & 0.6024 \\
        FCN 3197  & 0.6599 & 0.6639 & 0.7182 & 0.6469 & 0.7363 & 0.7504 & 0.7559 & \textbf{0.7635} \\
        INCEPTION 3197  & 0.7054 & 0.7257 & 0.7382 & 0.7312 & 0.7396 & 0.6764 & 0.6521 & 0.7209 \\
        \hline 
        MLP 2468  & 0.6149 & 0.6265 & 0.6423 & 0.6054 & 0.6701 & 0.6325 & 0.6436 & 0.6269 \\
        FCN 2468  & 0.7412  & 0.715  & 0.7452 & 0.6799 & 0.6113 & 0.733 & 0.7171 & 0.7252 \\
        INCEPTION 2468  & 0.7292 & 0.7434 & 0.7107 & 0.7268 & 0.7859 & \textbf{0.7912} & 0.7165 & 0.6913 \\
        \hline 
        MLP 2648  & 0.5958 & 0.6141 & 0.6044 & 0.5932 & 0.6193 & 0.6379 & 0.6728 & 0.6952 \\
        FCN 2648  & 0.7165 & 0.7279 & 0.7069 & 0.6899 & 0.7343 & 0.7223 & 0.7713 & 0.7619 \\
        INCEPTION 2648 & 0.7012 & 0.7226 & 0.7293 & 0.7717 & 0.7751 & \textbf{0.8342} & 0.8119 & 0.8166 \\
        \bottomrule
   \end{tabu}
}
\end{table}

Table~\ref{tab:cont_acc} summarizes the controller-based interactions, demonstrating that the InceptionTime classifier achieves the highest accuracy of 83.42\% for PIN 2648 at WS 100. The other classifiers show comparably high accuracies for the 2648 combination. PIN 2468 follows with an accuracy of 79.12\%. The PIN combinations 1379 and 3197 attain accuracies of approximately 76\%. The results generally exhibit improvements with increased WS, indicating that classifiers derive substantial benefits from longer sequences that encompass richer contextual motion information. 
These results are due to the motion requirements; PIN 2648 requires two diagonal complex spatial coordinates, from 2 to 6, and then 4 to 8. Novice users struggle with motion consistency, whereas experienced users perform more fluidly. In contrast, PIN 1379's simple left-to-right signal \textit{`Z'} diagonal motion aligns with natural motor habits, making it easier for all users and reducing classifier separability. This suggests that directional complexity enhances the detection of VR familiarity.

\setlength{\tabcolsep}{3pt}
\begin{table}[htb]
\centering
\caption{Hand-Tracking Classification Accuracies}
\label{tab:hand_acc}
\resizebox{\columnwidth}{!}{%
  
  \begin{tabu}{c || cccccccc}
  	\toprule
       WS & 50 & 60 & 70  & 80 & 90 & 100 & 110 & 120 \\ 
      	\hline \hline 
        MLP 1379  & 0.7562 & 0.7544 & 0.7763 & 0.7568 & 0.7339 & 0.7453 & 0.788 & 0.7992 \\
        FCN 1379  & 0.6702 & 0.669 & 0.7015 & 0.6616 & 0.6371 & 0.7045 & 0.7349 & \textbf{0.8043} \\
        INCEPTION 1379  & 0.7697 & 0.7549 & 0.6588 & 0.7244 & 0.5857 & 0.5975 & 0.6556 & 0.7139 \\
        \hline 
        MLP 3197  & 0.7232 & 0.6601 & 0.6698 & 0.7031 & 0.7583 & 0.7412 & 0.7055 & 0.6995  \\
        FCN 3197  & 0.7806 & 0.8265 & 0.8332 & 0.7859 & 0.8076 & 0.8285 & 0.8068 & 0.7798 \\
        INCEPTION 3197  & 0.8225 & 0.8506 & 0.8010 & 0.8948 & \textbf{0.9205} & 0.8838 & 0.9174 & 0.8975 \\
        \hline 
        MLP 2468  & 0.7451 & 0.7409 & 0.7623 & 0.7389 & 0.7628 & 0.6981 & 0.7433 & \textbf{0.7725} \\
        FCN 2468  & 0.6526 & 0.5869 & 0.6172 & 0.6809 & 0.6455 & 0.6339 & 0.6547 & 0.6082 \\
        INCEPTION 2468  & 0.6569 & 0.6997 & 0.6819 & 0.6224 & 0.5927 &0.6115  &0.6987  & 0.7604  \\
        \hline 
        MLP 2648  & 0.7068 & 0.7585 & 0.7855 & 0.8442 & 0.8339 & 0.8756 & 0.8830 & 0.8927 \\
        FCN 2648  & 0.7906 & 0.6979 & 0.7108 & 0.6411 & 0.6914 & 0.7523 & 0.7707 & 0.7117 \\
        INCEPTION 2648  & 0.8722 & 0.8082 & \textbf{0.9016} & 0.8369 &0.7601  & 0.7911 & 0.8763 & 0.8167  \\
        \bottomrule
   \end{tabu}
}
\end{table}

Table~\ref{tab:hand_acc} presents the results of the hand-tracking interactions, demonstrating significantly higher accuracies compared to controller interactions. The highest accuracy of 92.05\% is observed for PIN 3197 at a WS of 90. PIN 2648 shows the second-highest accuracy of 90.16\%, underscoring the premise that natural hand-tracking interactions provide more robust biometric signals capable of reliably differentiating between experienced and inexperienced users. Additionally, the accuracies of 80.43\% for PIN 1379 and 77.25\% for PIN 2468 further substantiate the effectiveness of hand-tracking data in detecting user familiarity. Users appear to gradually acclimate to the environment based on the order of entry: 1379, 2468, 2648 and 3197. Unlike other PINs, 3197 requires users to move from right to left using hand tracking. This movement may necessitate a period of re-acclimatization for users unfamiliar with VR.

\setlength{\tabcolsep}{3pt}
\begin{table}[htb]
\centering
\caption{Cross Device Classification Accuracies}
\label{tab:dev_acc}
\resizebox{\columnwidth}{!}{%
  \begin{tabu}{c || cccccccc}
  	\toprule
       WS & 50 & 60 & 70  & 80 & 90 & 100 & 110 & 120 \\ 
      	\hline \hline 
        MLP 1379  & 0.6418 & 0.6388 & 0.6548 & 0.6458 & 0.6224 & 0.6268 & 0.6199 &0.6155  \\
        FCN 1379  & 0.6700 & 0.6841 & 0.7068 & 0.6668 & 0.7009 & 0.7405 & \textbf{0.7516} & 0.7219  \\
        INCEPTION 1379  & 0.5976 & 0.6055 & 0.6194 & 0.6118 & 0.6033 & 0.6436 & 0.6424 & 0.6478 \\
        \hline 
        MLP 3197  & 0.5921 & 0.6178 & 0.6159 & 0.6408 & 0.6469 & 0.6599 & 0.6602 &0.6224  \\
        FCN 3197  & 0.5958 & 0.6077 & 0.5800 & 0.6516 & 0.6941 & 0.6898 & \textbf{0.7265} & 0.6965 \\
        INCEPTION 3197  & 0.5802 & 0.5762  & 0.5837 & 0.5979 & 0.6026 & 0.5762 & 0.6008 & 0.5969 \\
        \hline 
        MLP 2468  & 0.6435 & 0.6242 & 0.6045 & 0.6355 & 0.6444 & 0.6754 & 0.6548 & 0.6781 \\
        FCN 2468  & 0.7021 & 0.6734 & 0.7091 & 0.6964 & 0.6899 & 0.7001 & \textbf{0.7889} & 0.7219 \\
        INCEPTION 2468  &0.6347  & 0.5952 & 0.6364 & 0.6144 & 0.6370 & 0.6256 & 0.6686 & 0.6577 \\
        \hline 
        MLP 2648  & 0.5806 & 0.5606 & 0.5913 & 0.6087 & 0.6148 & 0.6412 & \textbf{0.6597} & 0.6454 \\
        FCN 2648  & 0.6351 & 0.5943 & 0.6199 & 0.618 & 0.6399 & 0.6537 & 0.5982 & 0.5877 \\
        INCEPTION 2468  & 0.6187 & 0.6057 & 0.5918 & 0.6227 & 0.5998 & 0.5961 & 0.5931 & 0.6165 \\
        \bottomrule
   \end{tabu}
}
\end{table}

The cross-device classification results in Table~\ref{tab:dev_acc} show a decrease in performance when evaluating the model trained on controller interactions using hand-tracking data, though the results remain promising. PIN 2468 at WS 110 achieves the highest accuracy 78.89\%, followed by PIN 1379 having 75\%. In contrast, PINs 2648 and 3197 exhibit accuracies close to chance. 
This variation in performance likely stems from differences in user hand motion dynamics resulting from the distinct interaction mechanics and ergonomic affordances of controller-based and hand-tracking input modalities.

\setlength{\tabcolsep}{3pt}
\begin{table}[htb]
\centering
\caption{Mixed Device Classification Accuracies}
\label{tab:mix_acc}
\resizebox{\columnwidth}{!}{%
  \scriptsize%
  
  \begin{tabu}{c || cccccccc}
  	\toprule
       WS & 50 & 60 & 70  & 80 & 90 & 100 & 110 & 120 \\ 
      	\hline \hline 
        MLP 1379  & 0.6654 & 0.6886 & 0.6568 & 0.6691 & 0.6353 & 0.6426 & 0.6015 & 0.5951 \\
        FCN 1379  & 0.7072 & 0.7369 & 0.7627 & 0.7435 & 0.7451 & 0.7678 & 0.8113 & 0.8202  \\
        INCEPTION 1379  & 0.6967 & 0.7488 & 0.7961 & 0.7849 & 0.8045 & 0.8136 & \textbf{0.8225} & 0.8209 \\
        \hline 
        MLP 3197  & 0.6551 & 0.7028 & 0.6919 & 0.7329 & 0.6381 & 0.6277 & 0.6078 & 0.5914 \\
        FCN 3197  & 0.7462 & 0.7372 & 0.7673 & 0.7467 & 0.8786 & 0.8296 & 0.8993 & \textbf{0.9419}  \\
        INCEPTION 3197  & 0.6723 & 0.6264 & 0.6464 & 0.7185 & 0.6967 & 0.7216 & 0.7375 & 0.7496 \\
        \hline 
        MLP 2468  & 0.6653 & 0.6477 & 0.6731 & 0.6483 & 0.6591 & 0.6554 & 0.6725 & 0.6544 \\
        FCN 2468  & 0.7153 & 0.7737 & 0.7656 & 0.8211  & 0.7718 & 0.8701  & 0.7131 & 0.7591 \\
        INCEPTION 2468  & 0.7606 & 0.8204 & 0.7508 & 0.7392 & 0.7542 & 0.7621 & 0.7872 & \textbf{0.8765} \\
        \hline 
        MLP 2648  & 0.5763 & 0.6004 & 0.6512 & 0.6982 & 0.6732 & \textbf{0.7169} & 0.6294 & 0.6842  \\
        FCN 2648  & 0.6952 & 0.6594 & 0.6715 & 0.5731 & 0.6373 & 0.5644 & 0.6173 & 0.6067 \\
        INCEPTION 2648  & 0.6075 & 0.6284 & 0.5566 & 0.5813 & 0.5428 & 0.5394 & 0.5273 & 0.6077 \\
        \bottomrule
   \end{tabu}
}
\end{table}

Table~\ref{tab:mix_acc} presents the results of the mixed-device classification. The highest accuracy of 94.19\% is obtained for PIN 3197 at WS 120 using the FCN classifier. Accuracies of 87.65\% and 82.25\% are observed for PIN 2468 and PIN 1379, respectively. Furthermore, PIN 2648 achieves a 71\% accuracy using an MLP classifier. The InceptionTime and FCN models consistently demonstrate robust performances across various PIN combinations. Similarly, Mathis et al.~\cite{mathis2020knowledge} reported that the FCN model surpasses other methodologies in the context of VR security. These findings suggest that the integration of data from both interaction modalities significantly enhances the classifier performance by capturing biometric features across diverse user interactions.

Our results demonstrate that hand movement patterns are a highly effective biometric for assessing user familiarity with VR. The high classification accuracies observed across various modalities, particularly in mixed-device scenarios, highlight the significant potential of behavior-based biometric systems for adaptive VR environments.

\section{Conclusion and Future Work}
We present a novel approach to automatic detection of user VR familiarity by analyzing hand motion behavior during a passcode-based task. Our experiments demonstrate that VR familiarity can be effectively detected using both hand-tracking and controller-based inputs, with high classification accuracies of 92.05\% and 83.42\%, respectively. The results show that our approach, utilizing state-of-the-art deep learning classifiers, performs well across various input modalities, even when tested across devices, achieving an accuracy of 94.19\% in the mixed-device scenario. 
Our work fills a gap in the VR literature by focusing on skill acquisition rather than traditional identity or security metrics. Our approach enables VR systems to infer user familiarity and adapt in real time. A VR training application detects novice users and switches to assisted mode by featuring tutorials. This adaptation minimizes frustration and improves the learning curve, thereby enhancing the overall user experience. Familiarity detection personalizes VR platform settings based on user experience, making immersive technologies more accessible.

 Our evaluation is predicated on short-term interactions within a controlled task environment; therefore, longitudinal studies are needed to determine the stability of familiarity detection as users gain experience, or when assessments are conducted on different days. We plan to expand our dataset to include a broader spectrum of VR experience levels quantified by factors such as the accumulated hours in VR. Future work will focus on engaging a more diverse user demographic and incorporating additional VR tasks to validate our familiarity detection across various age groups, backgrounds, and application contexts, thus enhancing generalizability and applicability. This expansion will enable a more fine-grained understanding of how behavioral patterns evolve with experience. We aim to incorporate head movement trajectories and eye-gaze dynamics as biometric features to improve the familiarity classification. Future work will involve analyzing more complex bimanual tasks for a comprehensive assessment of motor control and spatial interaction in VR. Furthermore, we will explore transfer learning techniques for cross-application and cross-platform personalization. We aim to integrate familiarity detection mechanisms into adaptive VR interfaces to support dynamic onboarding, user training, and error mitigation in real-world immersive environments. Such personalized adjustments would help reduce user errors and frustration, facilitate a seamless introduction to VR for novices, and enhance the overall quality of the experience in immersive environments.

\bibliographystyle{IEEEtran}
\bibliography{main}

\end{document}